\documentclass{article}

\makeatletter 
\@addtoreset{equation}{section}
\makeatother  

\usepackage{amssymb}
\usepackage{indentfirst}
\usepackage{setspace}
\usepackage{amsmath, mathtools}
\usepackage{mathrsfs}
\usepackage[colorlinks, linkcolor=black, anchorcolor=black, citecolor=black]{hyperref}
\usepackage{graphicx}
\usepackage{epstopdf}
\usepackage{tocloft}
\usepackage{booktabs}
\usepackage{bookmark}
\usepackage[toc,page,title,titletoc,header]{appendix}
\title{\textbf{{{Strong Cosmic Censorship in Charged de Sitter spacetime with Scalar Field Non-minimally Coupled to Curvature}}}}

\author{Hong Guo\thanks{E-mail:gh710105@gmail.com}, Hang Liu\thanks{E-mail:hangliu@sjtu.edu.cn}
\\
\textit{Department of Physics and Astronomy},
\\
\textit{Shanghai Jiao Tong University, Shanghai 200240, China}
\\
Xiao-Mei Kuang \thanks{E-mail:xmeikuang@yzu.edu.cn}, Bin Wang\thanks{E-mail:wang$\_$b@sjtu.edu.cn}
\\
\textit{Center for Gravitation and Cosmology, College of Physical Science and Technology,}
\\
\textit{Yangzhou University, Yangzhou 225009, China}
\\
\textit{School of Aeronautics and Astronautics}
\\
\textit{Shanghai Jiao Tong University, Shanghai 200240, China}}

\date{}
\begin{document}
\large
\maketitle

\begin{abstract}
We examine the stability and the strong cosmic censorship in the Reissner-Nordstrom-de Sitter (RN-dS) black hole by investigating the evolution of a scalar field non-minimally coupled to the curvature.  We find that when the coupling parameter is negative, the RN-dS black hole experiences instability. The instability disappears when the coupling parameter becomes non-negative. With the increase of the coupling parameter, the violation of the strong cosmic censorship occurs at a larger critical charge ratio. But such an increase of the critical charge is suppressed by the increase of the cosmological constant. Different from the minimal coupling situation, it is possible to accommodate $\beta\ge1$ in the near extremal black hole when the scalar field is non-minimally coupled to curvature.  The increase of the cosmological constant can allow $\beta\ge1$ to be satisfied for even smaller value of the coupling parameter. The existence of $\beta\ge1$ implies that the resulting curvature can continuously cross the Cauchy horizon.
\end{abstract}

\setcounter{page}{0}

\newpage
\section{Introduction}
It is well-known that the existence of the Cauchy horizon ($\mathcal{CH}$), i.e., the inner horizon of black holes implies the loss of determinism of the physics laws beyond the $\mathcal{CH}$. To rescue the determinism, Penrose proposed his famous Strong Cosmic Censorship (SCC) long ago. The SCC is based on the fact that the $\mathcal{CH}$ of realistic black holes formed dynamically in the asymptotically flat spacetime is inextendible due to the occurrences of the so-called mass-inflation induced by the blue shift amplification effect of the unavoidable time-dependent remnant fields propagating along the $\mathcal{CH}$. The presence of this scenario relies essentially on the inverse power law decay of the fields in the exterior region of asymptotically flat black hole spacetime, which cannot compete with the aforementioned blue shift amplification\cite{Hod1}.

However, when one considers the de Sitter ($\Lambda>0$) black holes, the situation we discussed above will change dramatically. Because in an asymptotically de Sitter spacetime, the perturbations outside the black holes decay instead exponentially in the form of  $ e^{-\alpha t}$, where $\alpha$ stands for the spectral gap, determined by the relation $\alpha=\mathrm{inf} \{\mathrm{-Im}(\omega_{i})\}$ over all possible quasi-normal modes (QNMs) $\omega_{i}$\cite{Cardoso1}. This kind of exponential decay behavior could be fast enough to make that the inner horizon singularity so weak that the spacetime metric would be extendible beyond the $\mathcal{CH}$ as a weak solution to the Einstein field equation\cite{Chris1}, leading eventually to the violation of the SCC. The fate of the SCC depends on the delicate competition between the exponential decay outside the black hole and the blue shift amplification along the $\mathcal{CH}$ in the interior region of black hole. In particular, it is found that the SCC will not be respected if the universal condition $\frac{\alpha}{\kappa_{-}}>\frac{1}{2}$ is satisfied, where the $\kappa_{-}$ is the surface gravity of the $\mathcal{CH}$. Whence we can assert that the SCC will be violated if and only if the condition $\beta\equiv-\frac{\mathrm{Im}\;\omega}{\kappa_{-}}>\frac{1}{2}$ is satisfied by \textit{all} QNMs. This amounts to saying that if there exists one mode which does not meet this condition, the SCC will be respected.

In order to examine the stability of the Reissner-Nordstrom--de Sitter(RN-dS) black holes, the QNMs have been investigated at various aspects, including the charged scalar field perturbation in four dimension\cite{Wang1}, and the gravitational perturbation in higher dimensions\cite{Konoplya1,Cardoso3}. In particular, due to the above relation between QNMs and SCC, the QNMs of RN-dS black holes have recently attracted resurgent attentions \cite{Hod1,Cardoso1,Dias3,Hod2,Cardoso2,Zhang1,Dias1,Dias2,Zhang2,Rahman,Gwak1,Liu,Gwak} since the pioneering work \cite{Cardoso1}. The linear neutral massless scalar perturbation is studied respectively in dimension $d=4$ \cite{Cardoso1}, and in higher dimensions \cite{Liu}. As a result, SCC is found to be violated when the black hole parameters are taken in the near extremal regime. Such a violation of SCC becomes more severe under the coupled electromagnetic and gravitational perturbations \cite{Dias3}. The discussions have also been generalized to the charged massless/massive scalar perturbation, and it has been shown that the SCC will be saved from being violated provided that the field is charged properly, but there is still a parameter regime in which the violation of SCC occurs \cite{Hod1,Cardoso2,Dias1,Zhang1}. The similar result has also been obtained for the Dirac field perturbation \cite{Zhang2,Destounis1}. In addition, The non-linear evolution of massless neutral scalar perturbation is considered in RNdS space and it turns out that the SCC might not be saved by such nonlinear effects \cite{Luna1}. On the other hand, the SCC in lukewarm RNdS and Martnez-Troncoso-Zanelli black hole spacetime under the non-minimally coupled massive scalar perturbation was investigated and it was shown that the validity of the SCC depends on the characteristics of the scalar field \cite{Gwak1}. Last but not least, although the Kerr-de Sitter black holes share many similarities with the RNdS ones, it is found that there is no violation of SCC for the linear perturbations \cite{Dias2}. Later on, no violation is further extended to higher dimensional Kerr-dS backgrounds for minimally coupled fields \cite{Rahman}, as well as the non-minimally coupled massive scalar field \cite{Gwak}.

It was argued that minimal coupling is of
limited value in the context of effective field theories, while non-minimally coupled interactions were appreciated in the general study in gravity  \cite{Weinberg}. As an alternative gravitation theory, scalar field non-minimally coupled to curvature was considered in \cite{AMENDOLA} and the coupling term in Einstein-Hilbert action allows for a suitable solution for the existence of inflation. The evolution of a scalar field coupled to curvature in topological black hole spacetimes was investigated in \cite{Bin}. So far all investigations on the SCC are limited to fields that are minimally coupled to curvature in RN-dS spacetime, it is natural to generalize the study to the non-minimal coupling scalar field. The work \cite{Gwak} studied the SCC and the  quasinormal resonance of non-minimal coupling scalar field in the Lukewarm Reissner-Nordstr\"{o}m-de Sitter black holes, however they set the charge and the mass of the RN-dS black holes to be equal to each other. In our present work, we will not impose any relations between black hole charge $Q$ and mass $M$, and we will disclose the general validity of the strong cosmic censorship in four-dimensional Reissner-Nordstr\"{o}m-de Sitter black hole against non-minimal coupling massless neutral scalar field perturbations. Considering that in \cite{Cardoso1} the SCC for massless neutral scalar perturbation with minimal coupling has been investigated, it is interesting to reveal the rich influences of the coupling parameter on the perturbation stability and the validity of the SCC.

The rest of the work is organized as follows. In Section 2, we give a brief and general introduction of the RN-dS spacetime in $d\geq4$ dimensions and derive the basic equations which control the motion of linear scalar perturbations. In Section 3, we analyze the stability of RN-dS black holes under non-minimally coupled linear scalar perturbations to disclose the impact of coupling constant on the stability of the RN-dS spacetime. In Section 4, we would like to investigate the validity of the SCC under scalar perturbations with different coupling constants, and we conclude with some discussions in the last section.

\section{Scalar perturbations with non-minimal coupling and the relation between QNMs and SCC}
In this section, we would like to give a general derivation of the equation of motion of the scalar field in $d\geq4$ dimensional RN-dS background.
The metric of RN-dS spacetime in d-dimension ($d\geq4$) is given by \cite{Chabab1}
\begin{equation}
ds^2=-f(r)dt^2+\frac{1}{f(r)} dr^2+r^2 d\Omega_{d-2}^2,
\end{equation}
where
\begin{align}
f(r)=1-\frac{m}{r^{d-3}}+\frac{q^2}{r^{2(d-3)}}-\frac{2\Lambda}{(d-2)(d-1)}r^2,
\end{align}
and
\begin{equation}
\Lambda=\frac{(d-2)(d-1)}{2L^2}, \quad d\Omega^2_{d-2}=d\chi_2^2+\prod_{i=2}^{d-2}sin^2\chi_i d\chi_{i+1}^2,
\end{equation}
in which $L$ is the cosmological radius associated with cosmological constant $\Lambda$, while the parameter $q$ and $m$ are related to the electric charge $Q$ and the ADM mass $M$ of the black hole as
\begin{equation}
M=\frac{d-2}{16\pi}\omega_{d-2}m,\quad Q=\frac{\sqrt{2(d-2)(d-3)}}{8\pi}\omega_{d-2}q,\quad \omega_{d}=\frac{2\pi^{\frac{d+1}{2}}}{\Gamma(\frac{d+1}{2})},
\end{equation}
with $\omega_{d}$ being the volume of the unit d-sphere. Furthermore, we have the electromagnetic field $F$ and gauge potential $A$
\begin{equation}
F_{ab}=(dA)_{ab},\quad A_{a}=-\sqrt{\frac{d-2}{2(d-3)}}\frac{q}{r^{d-3}}(dt)_{a}
\end{equation}
The action of the scalar field with a coupling constant $\lambda$ coupled with constant Ricci curvature $\mathcal{R}=2d/(d-2)\Lambda$ is
\begin{equation}
S_{\psi}=-\frac{1}{2}\int \mathbf{d}^{d}x \sqrt{-g}(\overline{D_{\nu}\psi} D^{\nu}\psi+(\mu^2+\lambda \mathcal{R})|\psi|^2),
\end{equation}
by which the equation of motion and energy-momentum tensor of the scalar field in the $d$-dimensional curved spacetime can be obtained as
\begin{gather}
(D^{\nu}D_{\nu}-(\mu^2+\lambda \mathcal{R}))\psi=0
\\
T_{\mu\nu}=\overline{D_{(\mu}\psi } D_{\nu)}\psi-\frac{1}{2}g_{\mu\nu}(\overline{D_{\rho}\psi}\partial^{\rho}\psi+\mu^2|\psi|^2)+\lambda(g_{\mu\nu}\Box-\nabla_\mu\nabla_\nu+
G_{\mu\nu})|\psi|^2\label{eq15},
\end{gather}
where $G_{\mu\nu}$ is the Einstein tensor, the operator $D_{\nu}=\nabla_{\nu}-ieA_{\nu}$ is the extended covariant derivative, while $e$ and $\mu$ are the electric charge and the mass of the field, respectively. We expand $\psi$ in the following way
\begin{equation}
\psi(r,t,\theta)=\sum_{lm}e^{-i\omega t}\frac{\phi(r)}{r^{\frac{d-2}{2}}}Y_{lm}(\chi)\label{eq3}.
\end{equation}
Whence the equation of motion reads
\begin{equation}
\begin{split}
0=f(r)f'(r)\phi'(r)+f^2(r)\phi''(r)+(B_1+B_2)\phi(r)\label{eq14},
\end{split}
\end{equation}
in which
\begin{align}
B_1=\frac{(d-2)e^2 q^2r^{2(3-d)}}{2(d-3)}-\frac{\sqrt{2}e q r^{3-d}}{(\frac{d-3}{d-2})^{\frac{1}{2}}}\omega+\omega^2=(\omega-\Phi(r))^2\\
B_2=f(r)\left(-\frac{l(d+l-3)}{r^2}-\mu^2-\frac{2d}{d-2}\Lambda\lambda-\frac{(d-2)f'(r)}{2r}\right)-f^2(r)\frac{(d-4)(d-2)}{4r^2},
\end{align}
and
\begin{equation}
\Phi(r)=\frac{q e}{\sqrt{\frac{2(d-3)}{d-2}}r^{d-3}}.
\end{equation}
In the present paper, we focus only on the massless neutral scalar perturbations in the four-dimensional RN-dS spacetime, which corresponds to $e=\mu=0$ and $d=4$. As a result, the equation can be simplified into
\begin{equation}
f(r)f'(r)\phi'(r)+f^2(r)\phi''(r)+(\omega^2-V_{eff})\phi(r)=0,\label{eq12}
\end{equation}
where
\begin{equation}
V_{eff}=f(r)\left(\frac{l(l+1)}{r^2}+4\Lambda\lambda+\frac{f'(r)}{r}\right)\label{eq9}
\end{equation}
 As one can see, the only effect of non-minimal coupling constant $\lambda$ is that the effective potential $V_{eff}$ will be affected by the term $4\Lambda\lambda$. Note that we are considering the case with a positive cosmological constant $\Lambda$, thus a positive $\lambda$ will increase the value of $V_{eff}$ while a negative $\lambda$ will decrease $V_{eff}$ such that $V_{eff}$ will have a wider and deeper negative region, leading to a potential instability of the black hole spacetime, as we shall discuss in Section 3.

By introducing the tortoise coordinate $dr_{\ast}=\frac{dr}{f(r)}$, we finally arrive at the master equation in the form of a Schrodinger equation
\begin{equation}
\frac{d^2\phi(r)}{dr_{\ast}^{2}}+(\omega^2-V_{eff})\phi(r)=0\label{eq7},
\end{equation}
which gives rise to the spectrum of QNMs once the boundary conditions are imposed as follows
\begin{equation}\phi(r)\approx e^{-i\omega r_{\ast}}, (r\rightarrow r_{+});\qquad \phi(r)\approx e^{i\omega r_{\ast}}, (r\rightarrow r_{c}).\label{eq4}
\end{equation}

To relate the QNMs to SCC, we may as well go to the ingoing coordinate $v=t+r_{\ast}$ outside the black hole. Because our QNMs behaving as $e^{-i \omega v}$ across the event horizon are analytic functions in this coordinate, we can analytically continue our QNMs solutions to the inside of the black hole. In the inside of the black hole, we can go back to the $t$ coordinate, where our analytically continued solutions behave as $e^{-i\omega t}$. One should note that generically we will have both the ingoing and outgoing modes at the $\mathcal{CH}$, i.e.,
\begin{equation}
\phi_{out}\sim e^{i\omega r_{\ast}},\quad \phi_{in}\sim e^{-i\omega r_{\ast}}, \quad r\rightarrow r_{-}.
\end{equation}
Taking into account that the ingoing coordinate $v$ is singular while the outgoing coordinate $u=t-r_{\ast}$ is well defined at the $\mathcal{CH}$, we would like to move onto the $u$ coordinate, where the two modes behave as
\begin{equation}
\phi_{out}\sim e^{-i\omega u},\quad \phi_{in}\sim e^{-i\omega u-2i\omega r_{\ast}}\sim e^{-i\omega u}(r-r_{-})^{\frac{i\omega}{\kappa_{-}}},r\rightarrow r_{-}.
\end{equation}
Obviously, the non-smoothness comes solely from the $\phi_{in}$ mode. the $\mathcal{CH}$ is extendible if and only if the energy-momentum tensor (\ref{eq15}) consisting of terms like $\psi^2, \partial_\mu\psi\partial_\nu\psi$ and $\nabla_\mu\nabla_\nu\psi^2$ of the scalar field should be integrable near the $\mathcal{CH}$, which amounts to saying that $\int(r-r_{-})^{2(p-1)}dr$ should be finite with $p=\frac{i\omega}{\kappa_{-}}$.
This requirement leads to $2[Re(p)-1]>-1$. Thus the criterion for the violation of the SCC is given by
\begin{equation}
\beta=-\frac{\mathrm{Im\; \omega}}{\kappa_{-}}>\frac{1}{2}
\end{equation}
for all QNMs. This means that if we can find some modes which meet the condition
\begin{equation}
\beta\le\frac{1}{2},
\end{equation}
then the SCC would be respected.

It is noteworthy that by this criterion the SCC can still be violated even in the presence of the blow up of the curvature near the $\mathcal{CH}$, which becomes finite when $\beta\ge1$ for all the modes.

\section{Stability analysis}
In this section, we would like to conduct a stability analysis of the RN-dS black hole under the neutral massless non-minimally coupled scalar fields. By inspecting the effective potential (\ref{eq9}), we can see that the effective potential is affected by the coupling constant $\lambda$ in the form of $4\Lambda \lambda$. Whence we can infer that the instability may occur if $\lambda$ is negative with big enough magnitude no matter which angular number $l$ we choose. On the contrary, note that the RN-dS black hole is stable under the neutral massless minimally coupled($\lambda=0$) scalar field \cite{Wang1}, and the appearance of the positive coupling constant will increase the effective potential such that the stability of the black hole is reinforced.

\begin{figure}[thbp]
\center{
\includegraphics[scale=0.9]{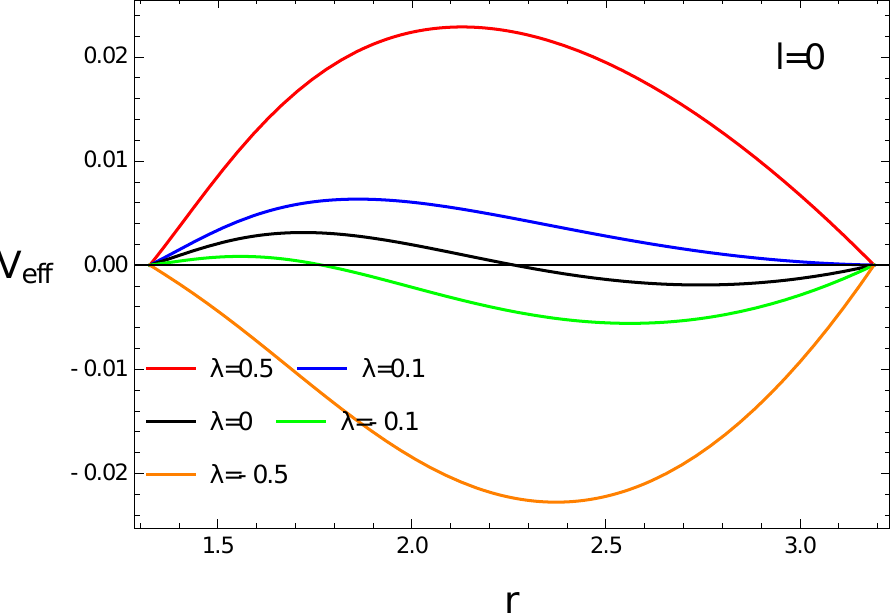}
\includegraphics[scale=0.9]{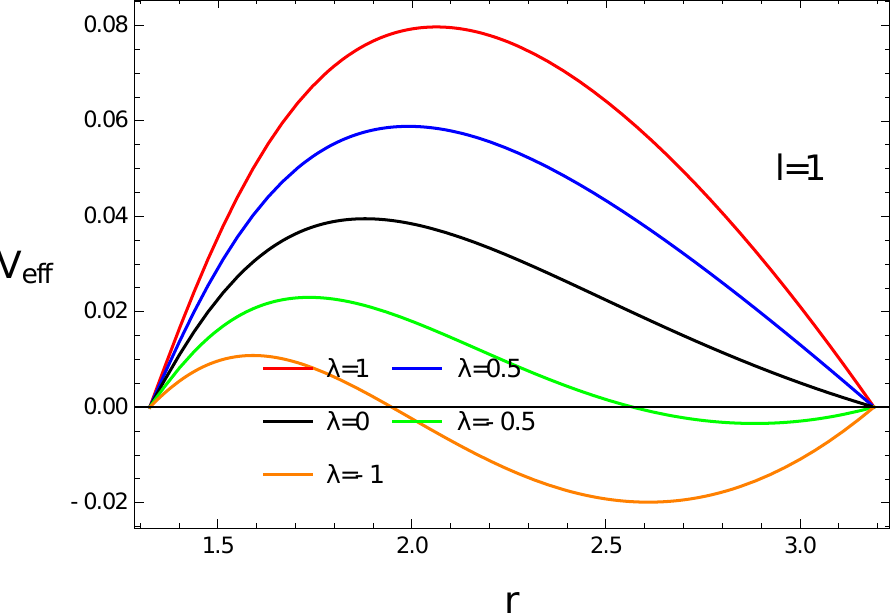}
\caption{$M=1,\Lambda=0.14,Q/Q_{max}=0.99$}.\label{fig1}}
\end{figure}

In Fig.\ref{fig1}, we show the behavior of effective potential $V_{eff}$ between the event horizon $r_+$ and the cosmological horizon $r_c$ for $l=0$ and $l=1$. As expected from the formula (\ref{eq9}), the larger angular number $l$, the more negative coupling constant is needed to have the occurrence of negative effective potential $V_{eff}$ in some region.  Such an occurrence signals the possible instability of the black hole\cite{Bronnikov,Ching}. To see what really happens, one is required to calculate the corresponding QNMs, where the positive imaginary part of QNM frequency $\omega=\omega_R+i\omega_I$ corresponds to the growing mode, indicating the instability of the black hole.

\begin{table}[!htbp]
\centering
\begin{tabular}{cccccc}
\toprule
angular number& $\lambda=-0.5$ & $\lambda=-0.1$  &$\lambda=-0.001$\\
\midrule
$l=0$&$0+i0.111107$&$0+i0.035139$&$0+i0.000470$\\
$l=1$&$0-i0.007209$&$0.181861-i0.047444$&$0.191424-i0.046261$\\
$l=5$&$0.730575-i0.045157$&$0.740796-i0.044912$&$0.743318-i0.044854$\\
$l=10$&$1.41688-i0.044862$&$1.42221-i0.044798$&$1.42352-i0.044782$\\
\bottomrule
\end{tabular}
\caption{ The QNMs for negative coupling constant when $M=1,\Lambda=0.14,Q/Q_{max}=0.99,n=0$.\label{table1}}
\end{table}

\begin{table}[!htbp]
\centering
\begin{tabular}{cccccc}
\toprule
angular number&$\lambda=0$ & $\lambda=0.1$&$\lambda=0.5$\\
\midrule
$l=0$&$0.041017-i0.069190$ &$0.057096-i0.049195$&$0.143892-i0.043464$\\
$l=1$&$0.19152-i0.046245$ &$0.201098-i0.045302$&$0.237539-i0.043293$\\
$l=5$&$0.743344-i0.044854$ &$0.745888-i0.044796$&$0.756032-i0.044575$\\
$l=10$&$1.42354-i0.044782$&$1.42487-i0.044766$&$1.43019-i0.044704$\\
\bottomrule
\end{tabular}
\caption{ The QNMs for positive coupling constant when $M=1,\Lambda=0.14,Q/Q_{max}=0.99,n=0$.\label{table2}}
\end{table}

To this end, we calculate the QNM frequency $\omega$ numerically by using the asymptotic iteration method (AIM)\cite{Cho1}. As such, we introduce a new variable $\xi$
\begin{equation}
\xi=\frac{1}{r}\label{eq5}.
\end{equation}
 rewrite $\phi(r)$ in terms of the new variable $\xi$ as follows
\begin{equation}
\phi(\xi)=(\xi_+-\xi)^{-\frac{i\omega}{2\kappa_{+}}}(\xi-\xi_c)^{-\frac{i\omega}{2\kappa_{c}}}\chi(\xi)\label{eq6}
\end{equation}
with $\xi_+ =r_+^{-1}$ and $\xi_c=r_c^{-1}$, and recast Eq.(\ref{eq12}) into the form as
\begin{equation}
\chi''(\xi)=\lambda_{0}(\xi) \chi'(\xi)+s_{0}(\xi)\chi(\xi)
\end{equation}

We list the $n=0$ QNMs with the angular number $l=0,1,5,10$ in Table \ref{table1} and Table \ref{table2} for negative and non-negative $\lambda$, respectively. For the negative $\lambda$, one can see there always exists a purely imaginary unstable mode at $l=0$. Moreover, the less negative the coupling constant $\lambda$ is, the smaller the imaginary of part of this unstable mode is. While for $\lambda\ge0$, the unstable mode does not exist, indicating that the black hole is stable.

\section{Strong cosmic censorship}

\subsection{The fate of SCC}

\begin{figure}[thbp]
\center{
\includegraphics[scale=0.35]{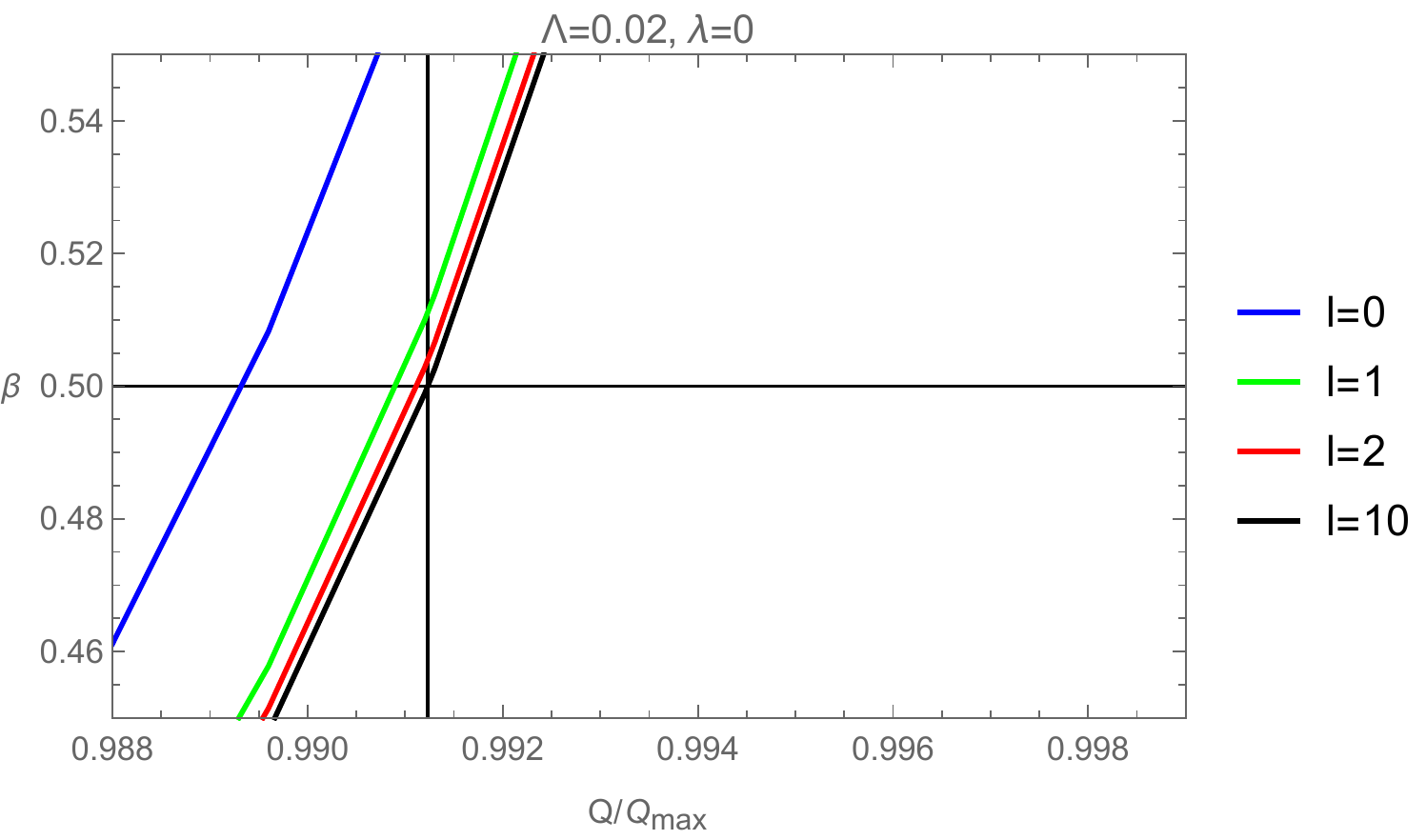}
\includegraphics[scale=0.35]{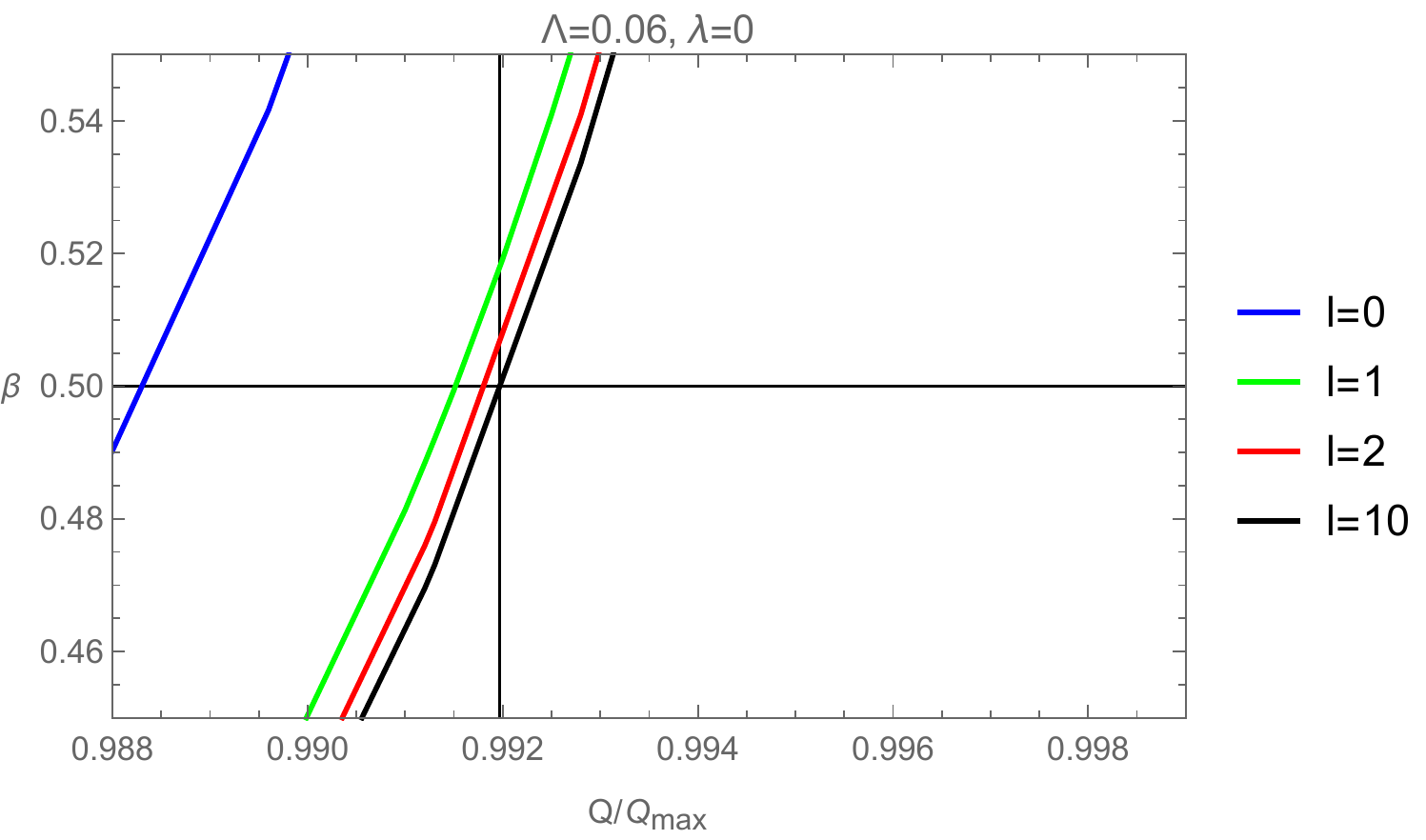}
\includegraphics[scale=0.35]{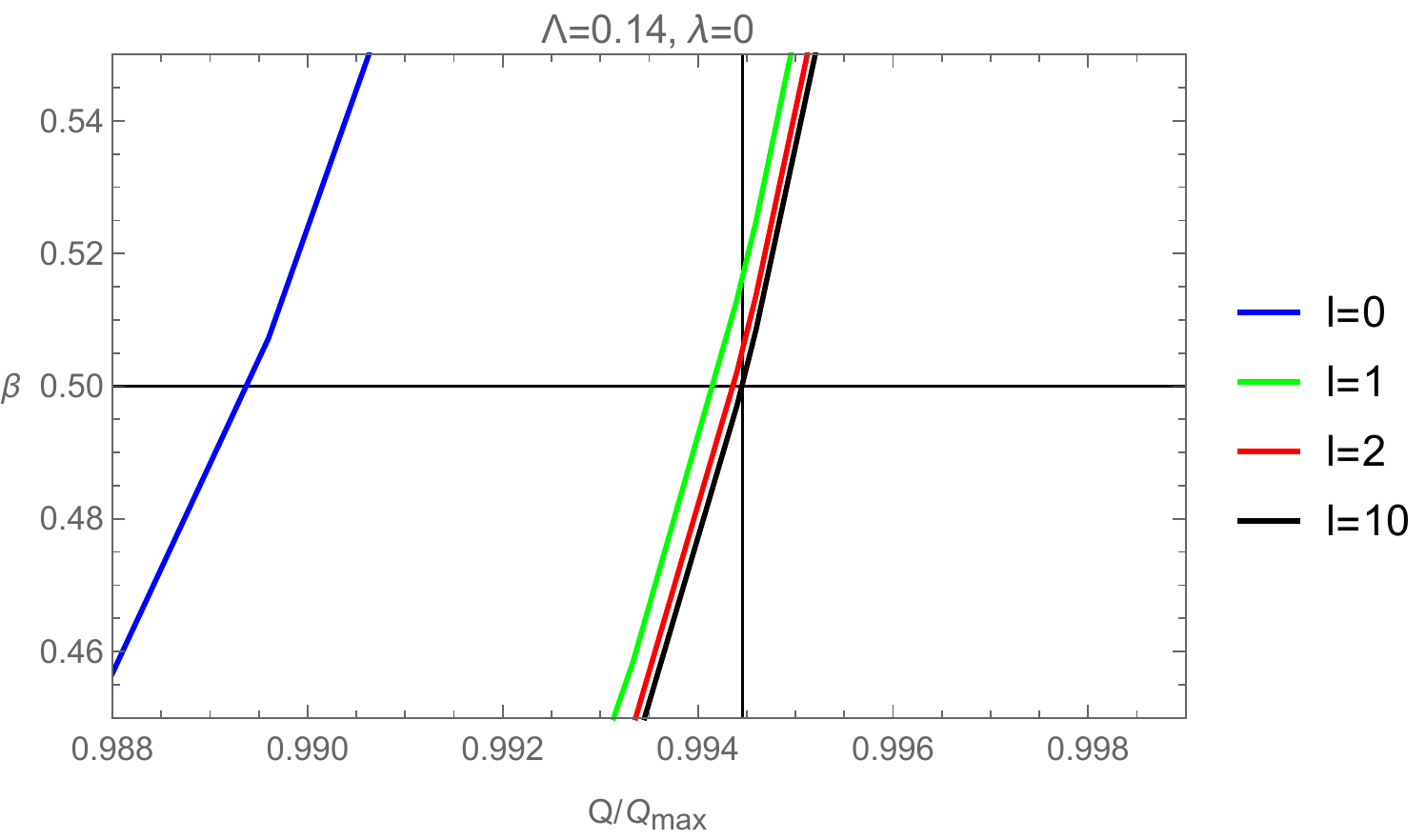}
\includegraphics[scale=0.35]{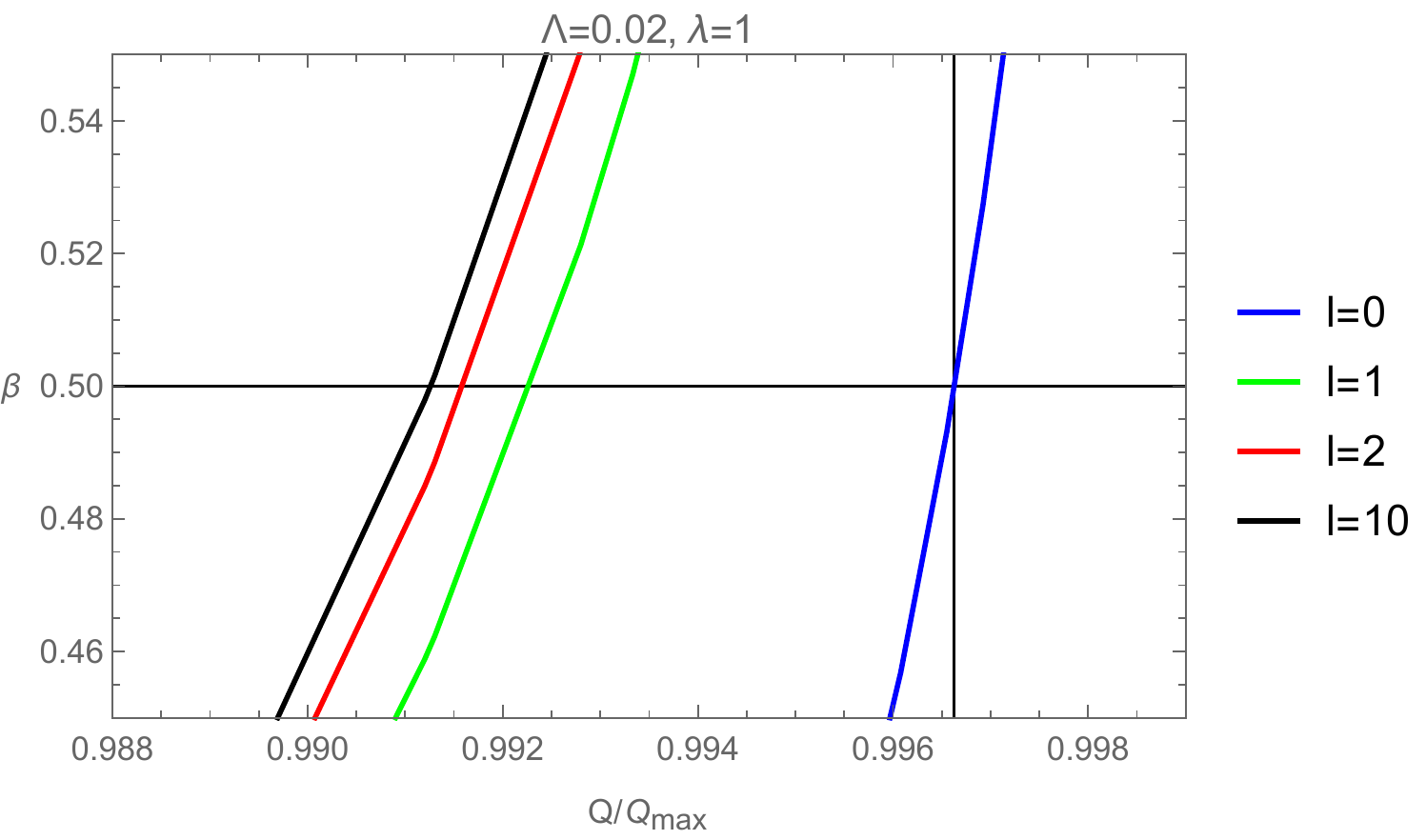}
\includegraphics[scale=0.35]{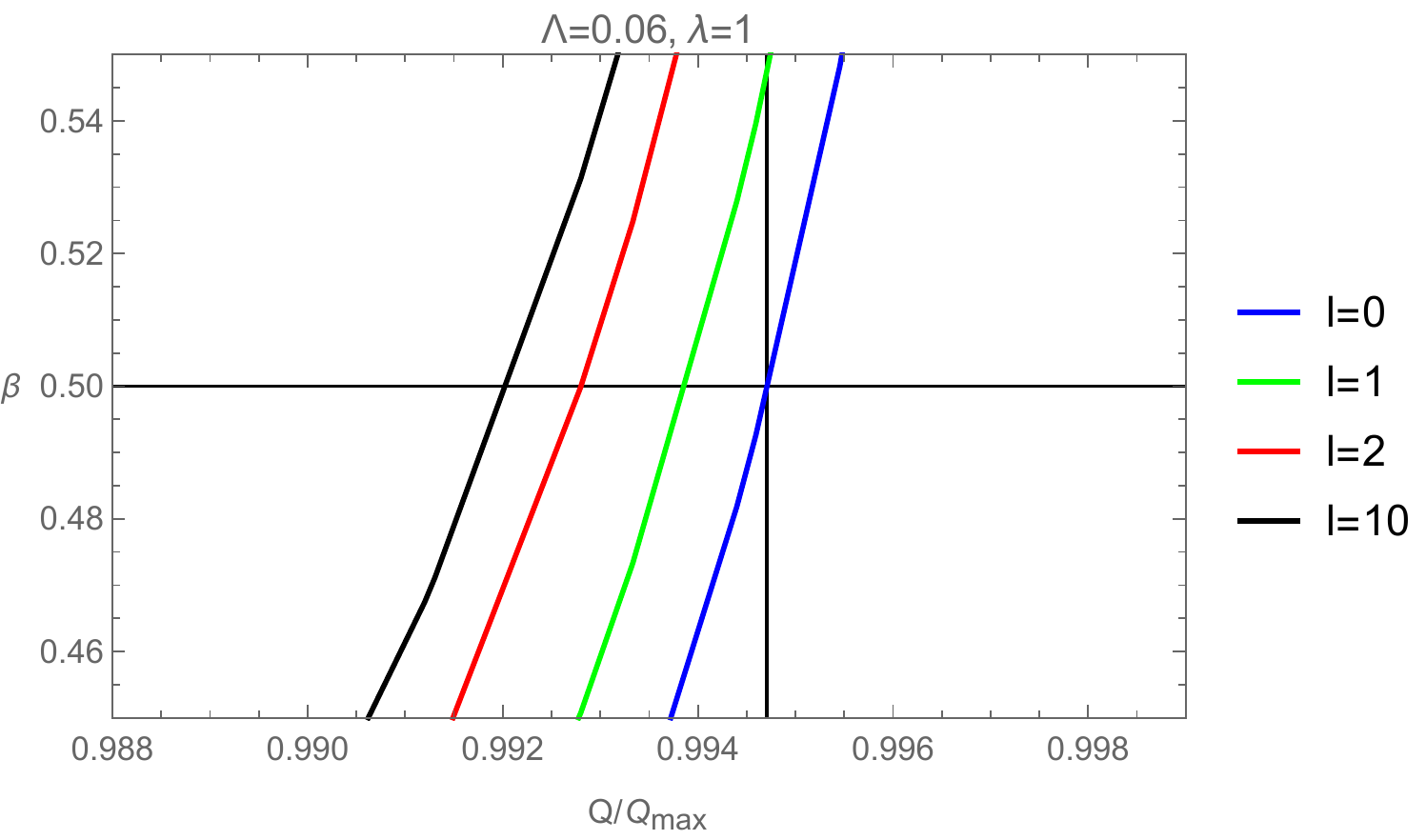}
\includegraphics[scale=0.35]{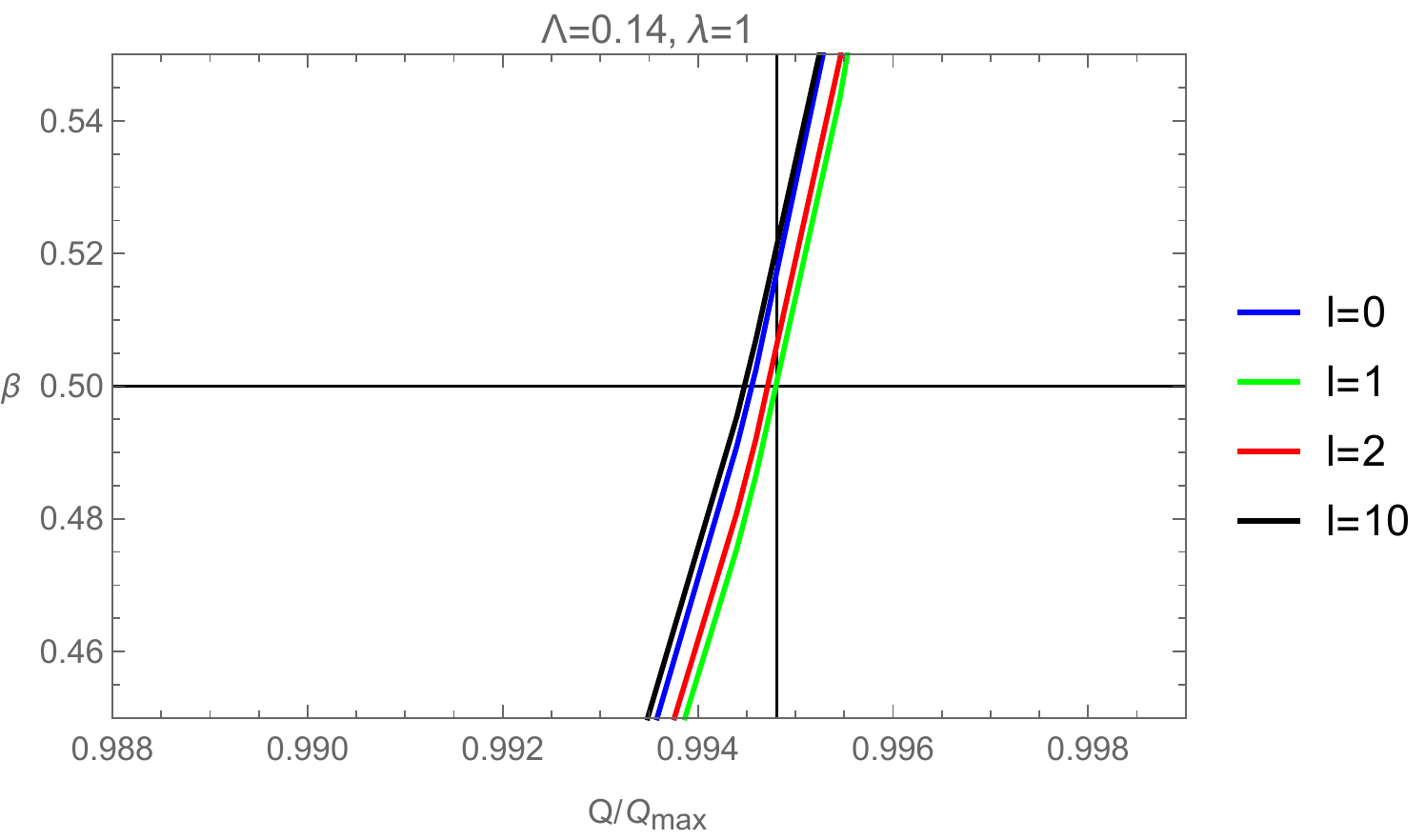}
\includegraphics[scale=0.35]{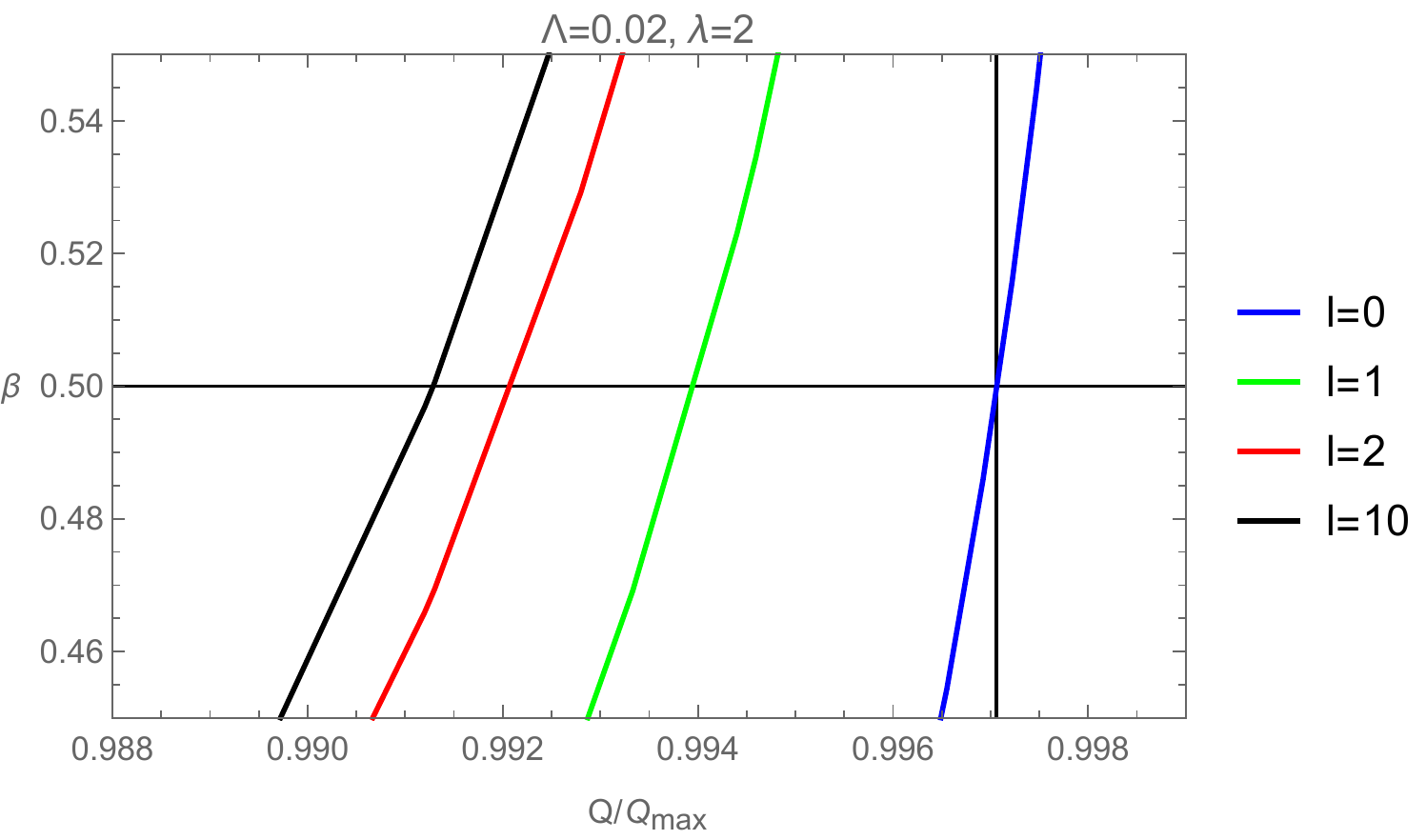}
\includegraphics[scale=0.35]{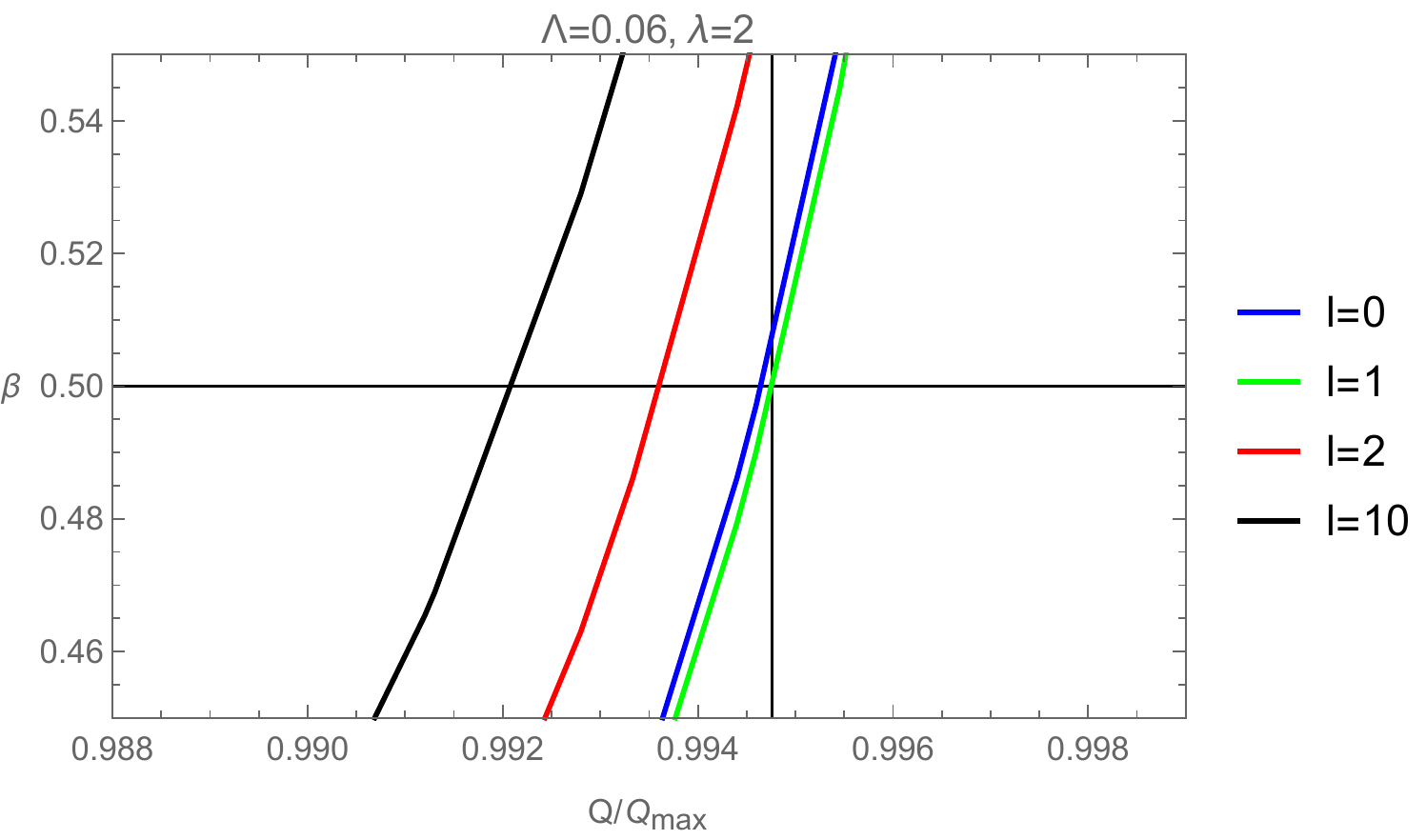}
\includegraphics[scale=0.35]{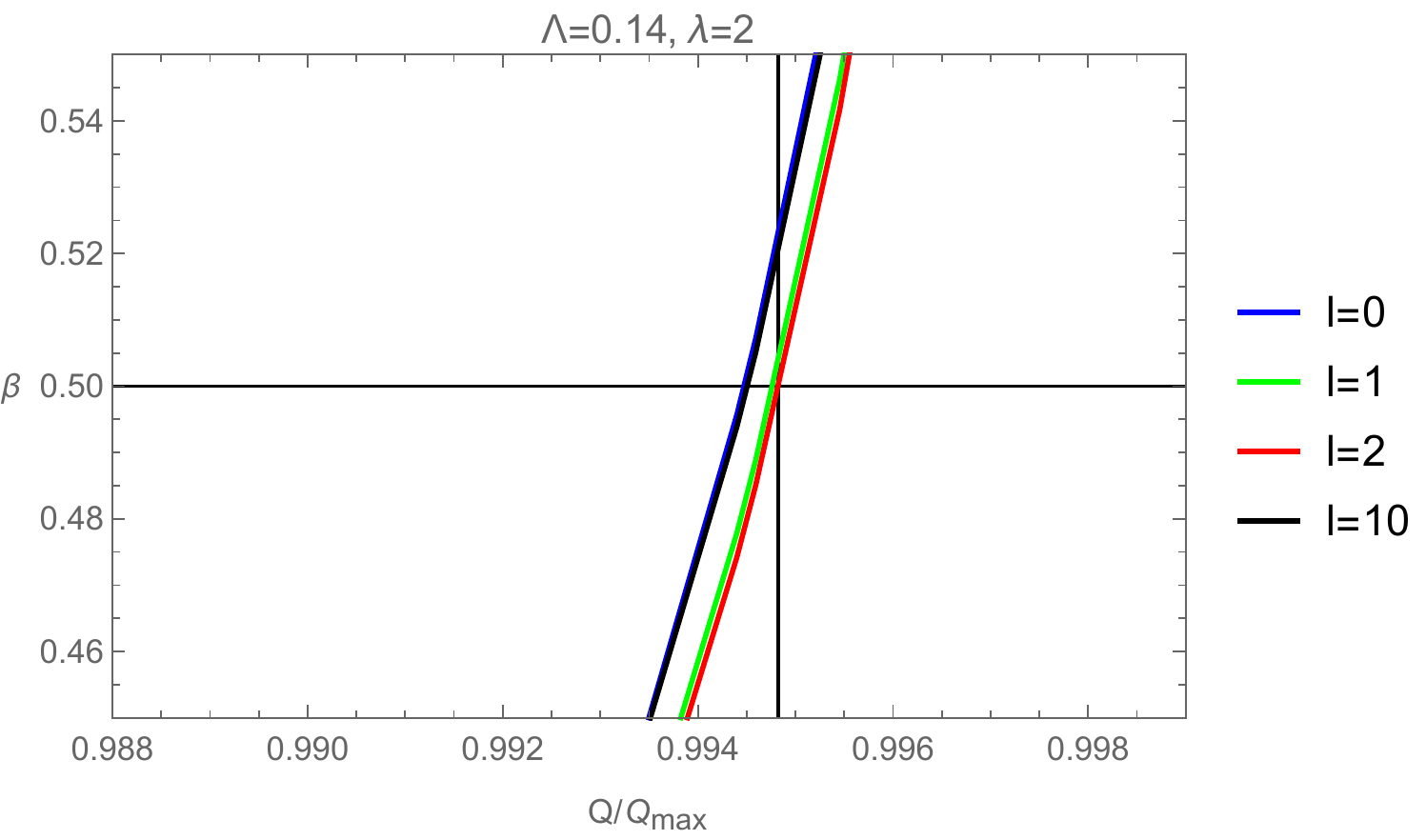}
\caption{\label{l1} The figures show how the different dominant modes cross through the horizontal black line $\beta=0.5$ in the different coupling parameters $\lambda=0,1,2$ when we fix the cosmological constant $\Lambda=0.02,0.06,0.14$. The critical charge ratio $\frac{Q_c}{Q_{max}}$ is determined by the intersection of the black vertical line with the horizontal line. }}
\end{figure}

Note that we always have an unstable mode for the negative coupling parameter, so regarding the potential violation of the SCC, below we shall focus only on the non-negative coupling parameter. We plot the most dominant modes for $l=0,1,2,10$ in Fig.\ref{l1}, where each column corresponds to the result for the same cosmological constant with the coupling parameter increased from the top to the bottom, and the black vertical line is used to indicate the critical charge ratio $\frac{Q_c}{Q_{max}}$ for the violation of the SCC.

\begin{figure}[thbp]
\center{
\includegraphics[scale=0.6]{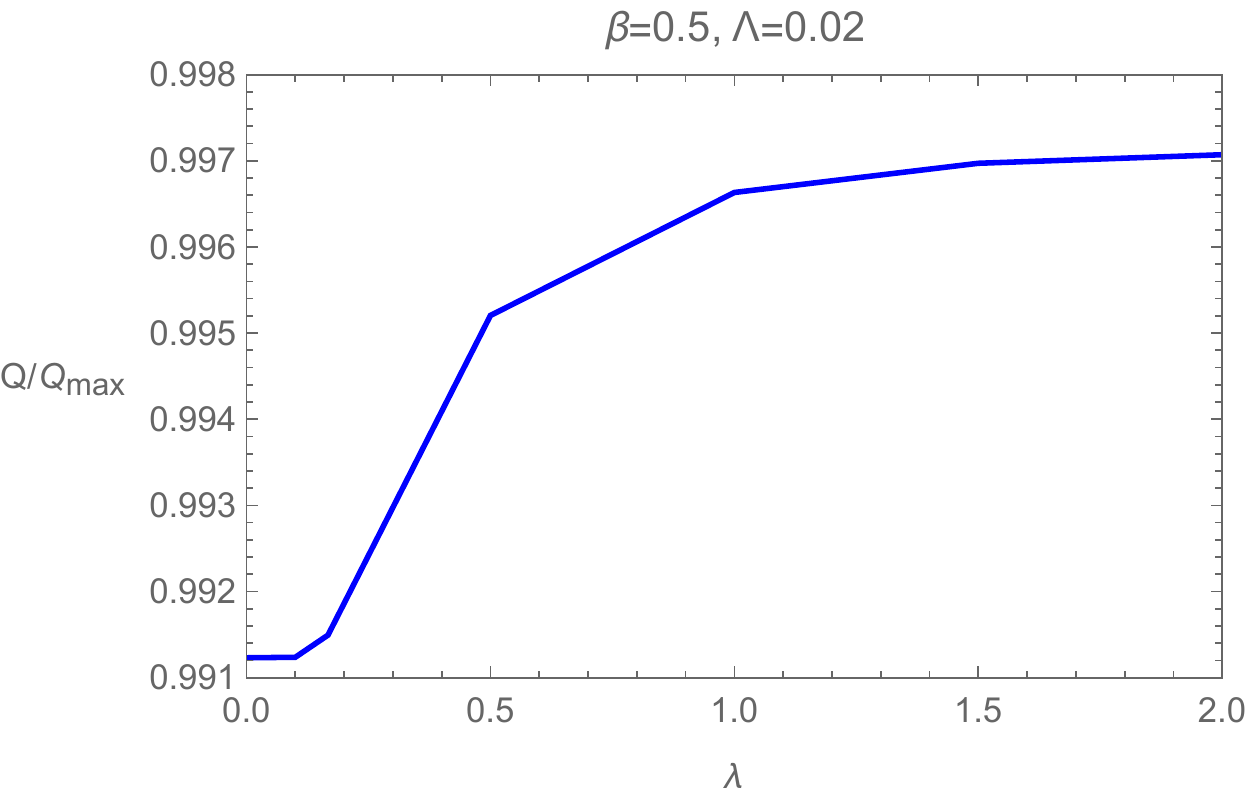}
\includegraphics[scale=0.6]{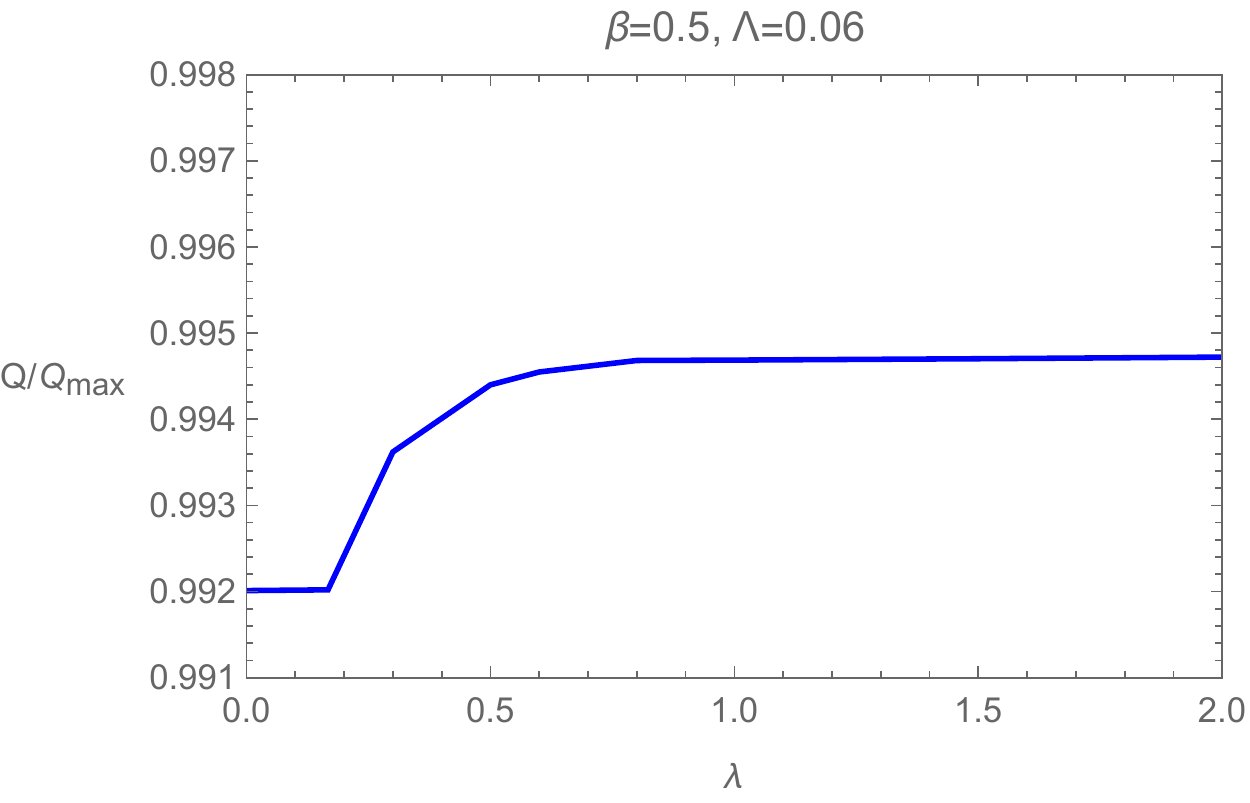}
\includegraphics[scale=0.6]{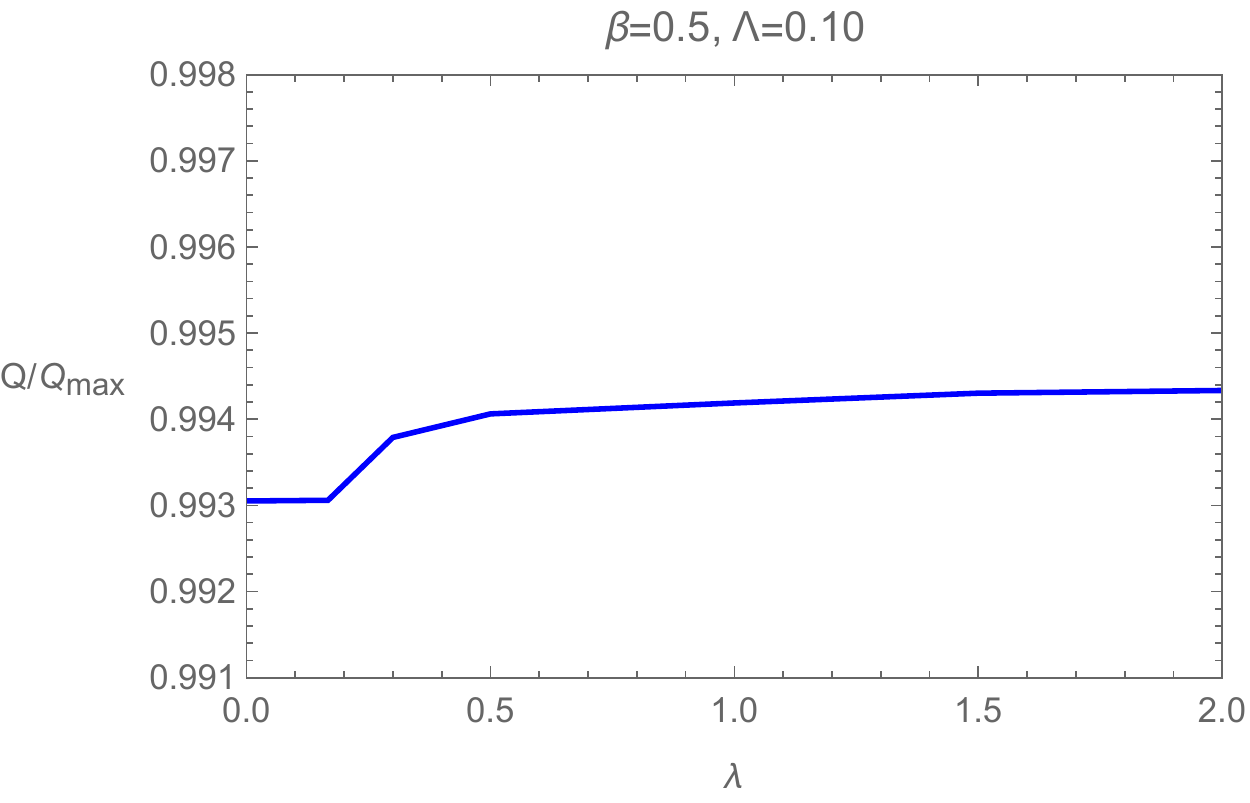}
\includegraphics[scale=0.6]{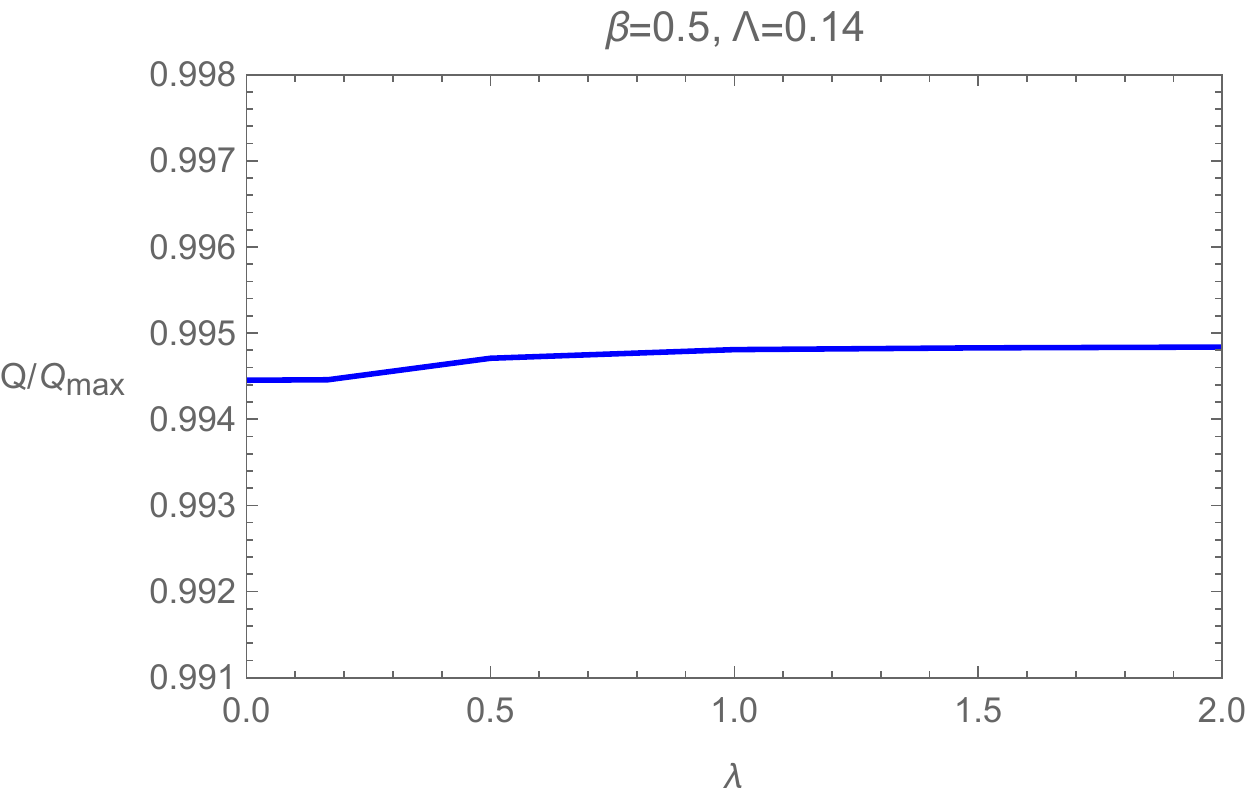}
\caption{\label{l2} The variation of the critical charge ratio $\frac{Q_c}{Q_{max}}$ with the increase of the coupling parameter.  }}
\end{figure}

As one can see, the violation of the SCC always occurs no matter how we choose the value of the coupling parameter and cosmological constant. In addition, for a fixed cosmological constant, the aforementioned critical charge ratio increases with the coupling parameter. So the larger the coupling parameter is, the harder the violation of the SCC is. To scrutinize such a behavior, we plot the variation of the critical charge ratio as the increase of the coupling parameter in Fig.\ref{l2}. First, it seems that such an increase will be saturated at a large coupling parameter. That amounts to saying that a larger coupling parameter will not affect the critical charge for the violation of the SCC any more. Second, the larger the cosmological constant is, the tinier such an increase is. In this sense, the cosmological constant seems to play a role in refraining the effect of the coupling parameter onto the critical charge.


\subsection{The blow up of the curvature}

It has been shown in \cite{Dias3} that $\beta<1$ for all the modes in the case of the minimal coupling, so the curvature blows up although the $\mathcal{CH}$ is extendible. Note that our scalar is coupled with the curvature, thus it is interesting to ask whether the curvature can cross the $\mathcal{CH}$ continuously when the coupling parameter is non-vanishing.

\begin{figure}[thbp]
\center{
\includegraphics[scale=0.5]{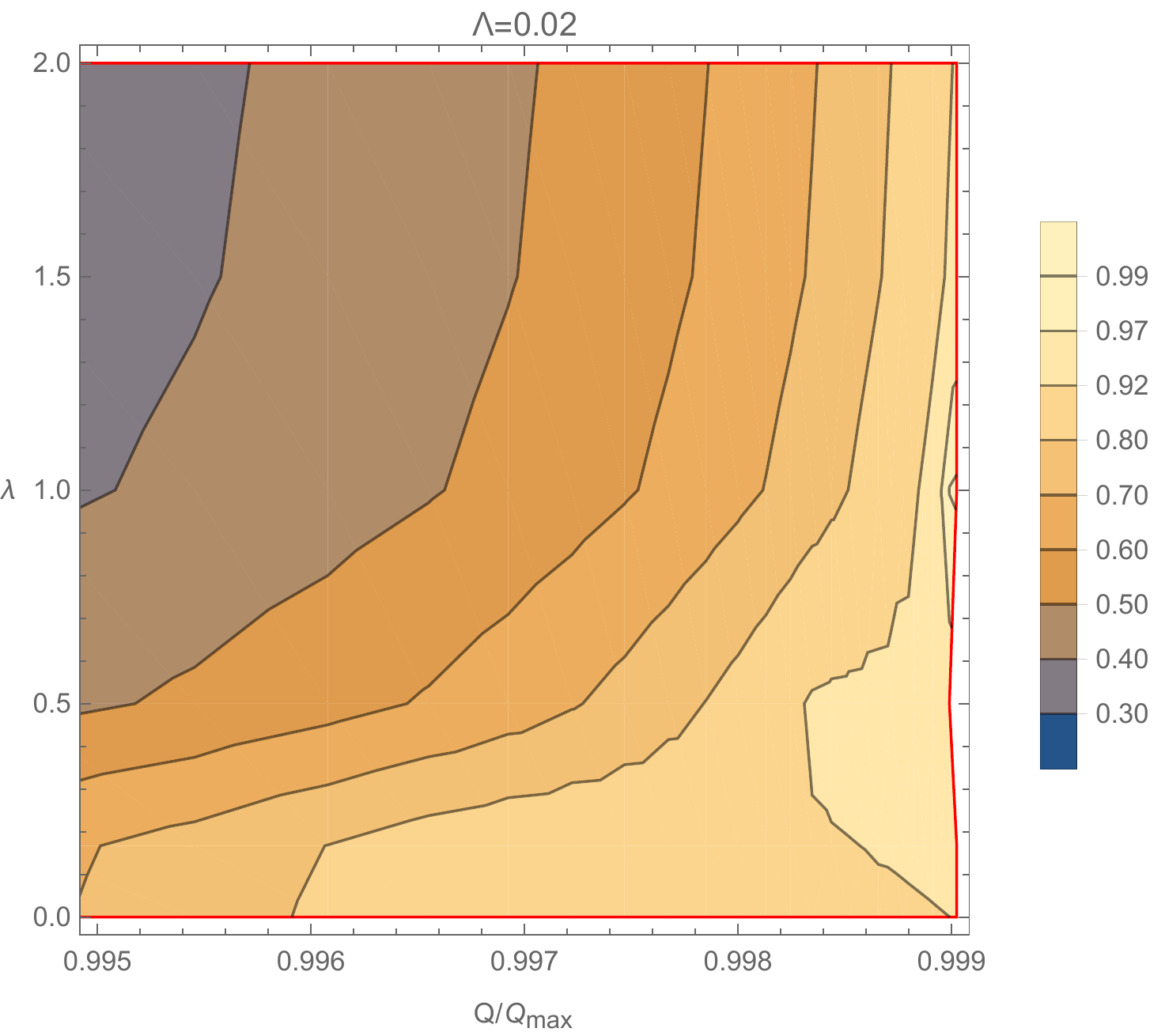}
\includegraphics[scale=0.5]{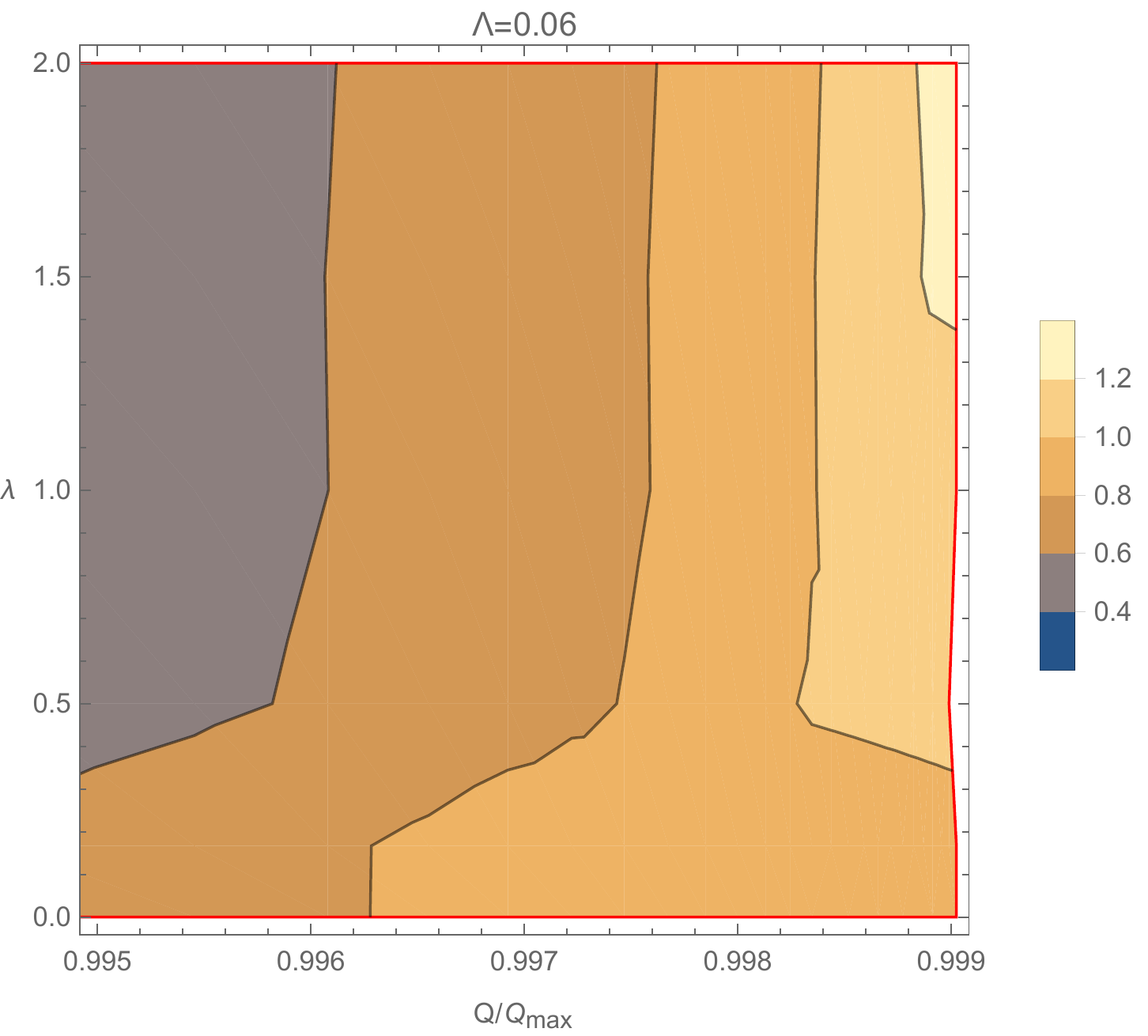}
\includegraphics[scale=0.5]{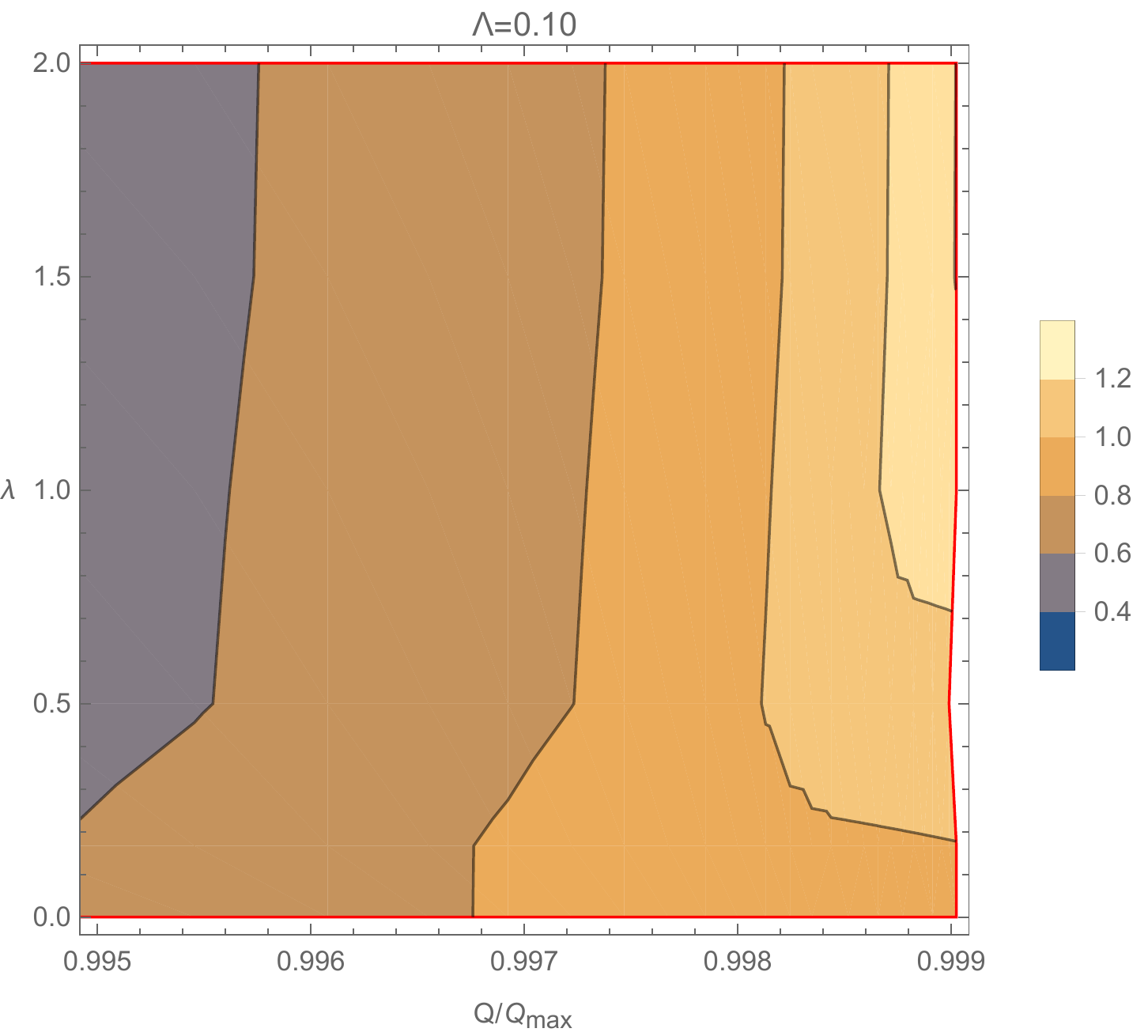}
\includegraphics[scale=0.5]{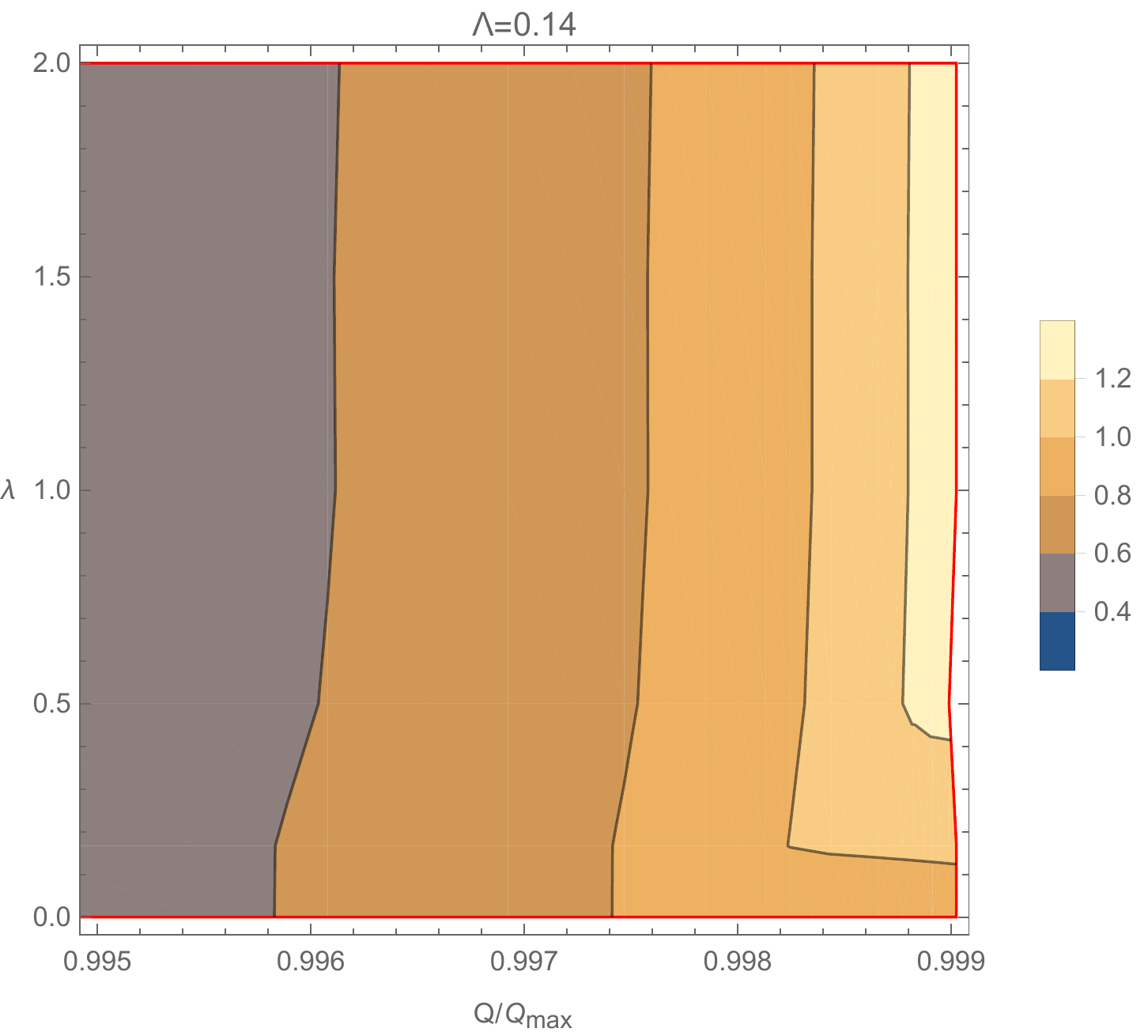}
\caption{\label{l4} The contour of $\beta$ for the most dominant mode in the space of the charge ratio $\frac{Q}{Q_{max}}$ and coupling parameter $\lambda$. }}
\end{figure}

As such, we list the most dominant QNM data at each angular number in Table \ref{t1} to \ref{t3}  for the charge ratio $\frac{Q}{Q_{max}}=0.9990244$, where the most dominant mode is blackened. As we see, $\beta\ge1$ can happen to all modes when the coupling parameter is non-vanishing. Furthermore, in Fig.\ref{l4}, we plot the contour of $\beta$ for the most dominant mode in the space of the charge ratio $\frac{Q}{Q_{max}}$ and coupling parameter $\lambda$. As one can see, when the cosmological constant is small, $\beta$ will not go beyond the limit $\beta=1$. When the cosmological constant is increased, the region of $\beta\ge1$ emerges in the near extremal regime and the corresponding range of coupling parameter is enlarged toward the small values.

\begin{table}
\centering
\begin{tabular}{|l|c|c|c|}
\hline
$\frac{Im(\omega)}{\kappa_{-}}$ & $\lambda=0$ & $\lambda=1$ & $\lambda=2$\\ \hline
$l=00$ & \textbf{0.0 -$i$0.910296} &  \textbf{0.0 -$i$1.135366} & \textbf{8.520377 -$i$1.341330}\\ \hline
$l=01$ & $7.313771 -i1.777196$ & $8.904893 -i1.452243$ & $10.538373 -i1.319612$\\ \hline
$l=02$ & $12.363690 -i1.731134$ & $13.289741 -i1.605510$ & $14.250843 -i1.490133$\\ \hline
$l=03$ & $17.378690 -i1.718803$ & $18.036803 -i1.655218$ & $18.706578 -i1.592246$\\ \hline
$l=04$ & $22.381133 -i1.713772$ & $22.892211 -i1.675494$ & $23.408555 -i1.637223$\\ \hline
$l=05$ & $27.377763 -i1.711241$ & $27.795629 -i1.685680$ & $28.216331 -i1.660093$\\ \hline
$l=06$ & $32.371227 -i1.709786$ & $32.724677 -i1.691516$ & $33.080409 -i1.673191$\\ \hline
$l=07$ & $37.362790 -i1.708877$ & $37.669058 -i1.695166$ & $37.976397 -i1.681435$\\ \hline
$l=08$ & $42.353112 -i1.708271$ & $42.623293 -i1.696914$ & $42.894239 -i1.686921$\\ \hline
$l=09$ & $47.342596 -i1.707848$ & $47.584302 -i1.699309$ & $47.826571 -i1.690759$\\ \hline
$l=10$ & $52.331439 -i1.707537$ & $52.549653 -i1.700606$ & $52.769209 -i1.693527$\\ \hline
\end{tabular}
\caption{\label{t1}The most dominant QNMs at each angular number in the case of $\Lambda=0.06,\frac{Q}{Q_{max}}=0.9990244$ for $\lambda=0,1,2$. }
\end{table}

\begin{table}
\centering
\begin{tabular}{|l|c|c|c|}
\hline
$\frac{Im(\omega)}{\kappa_{-}}$ & $\lambda=0$ & $\lambda=1$ & $\lambda=2$\\ \hline
$l=00$ & \textbf{0.0 -$i$0.894804} & \textbf{0.0 -$i$1.2941} & $9.490167-i1.467466$\\ \hline
$l=01$ & $6.883138 -i1.653291$ & $9.2443874 -i1.429376$ & \textbf{11.332765 -$i$1.402593}\\ \hline
$l=02$ & $11.792684 -i1.606922$ & $13.215042 -i1.497977$ & $14.618450 -i1.434143$\\ \hline
$l=03$ & $16.632350 -i1.595061$ & $17.644728 -i1.535904$ & $18.656848 -i1.487504$\\ \hline
$l=04$ & $21.449082 -i1.590290$ & $22.235217 -i1.553812$ & $23.022197 -i1.520846$\\ \hline
$l=05$ & $26.255466 -i1.587899$ & $26.898147 -i1.563277$ & $27.541511 -i1.540116$\\ \hline
$l=06$ & $31.056298 -i1.586531$ & $31.599857 -i1.568825$ & $32.143897 -i1.551832$\\ \hline
$l=07$ & $35.853795 -i1.585678$ & $36.324751 -i1.572329$ & $36.796042 -i1.559400$\\ \hline
$l=08$ & $40.649209 -i1.585116$ & $41.064541 -i1.574717$ & $41.480342 -i1.564540$\\ \hline
$l=09$ & $45.442989 -i1.584706$ & $45.814732 -i1.576372$ & $46.186605 -i1.568186$\\ \hline
$l=10$ & $50.235561 -i1.584247$ & $50.572279 -i1.577484$ & $50.908526 -i1.570854$\\ \hline
\end{tabular}
\caption{\label{t2}The most dominant QNMs at each angular number in the case of $\Lambda=0.10,\frac{Q}{Q_{max}}=0.9990244$ for $\lambda=0,1,2$.}
\end{table}

\begin{table}
\centering
\begin{tabular}{|l|c|c|c|}
\hline
$\frac{Im(\omega)}{\kappa_{-}}$ & $\lambda=0$ & $\lambda=1$ & $\lambda=2$\\ \hline
$l=00$ & \textbf{0.0 -$i$0.865473} & $6.692527 -i1.401811$ & $9.562236 -i1.415554$\\ \hline
$l=01$ & $6.295308 -i1.445910$ & \textbf{9.046770 -$i$1.349113} & $11.273336 -i1.358708$\\ \hline
$l=02$ & $10.917835 -i1.415302$ & $12.637447 -i1.360230$ & \textbf{14.238852 -$i$1.343550}\\ \hline
$l=03$ & $15.444360 -i1.407717$ & $16.682366 -i1.374939$ & $17.882357 -i1.355230$\\ \hline
$l=04$ & $19.940636 -i1.404689$ & $20.906034 -i1.383618$ & $21.855045 -i1.367681$\\ \hline
$l=05$ & $24.423416 -i1.403178$ & $25.214220 -i1.388650$ & $25.996507 -i1.376473$\\ \hline
$l=06$ & $28.898990 -i1.402315$ & $29.568563 -i1.391740$ & $30.233139 -i1.382379$\\ \hline
$l=07$ & $33.370262 -i1.401774$ & $33.950764 -i1.393755$ & $34.528132 -i1.386401$\\ \hline
$l=08$ & $37.838760 -i1.401415$ & $38.351116 -i1.395128$ & $38.861309 -i1.389234$\\ \hline
$l=09$ & $42.305368 -i1.401164$ & $42.763880 -i1.396114$ & $43.220798 -i1.391338$\\ \hline
$l=10$ & $46.770491 -i1.401242$ & $47.185146 -i1.396836$ & $47.599524 -i1.393582$\\ \hline
\end{tabular}
\caption{\label{t3}The most dominant QNMs in the case of $\Lambda=0.14,\frac{Q}{Q_{max}}=0.9990244$ for $\lambda=0,1,2$.}
\end{table}

\section{Conclusions}

In this paper, we consider the perturbation of the non-minimally coupled scalar field in four dimensional RN-dS spacetime. In particular, we have investigated the stability of the RN-dS black hole and the SCC under such a perturbation by applying AIM method to calculate out the QNMs. When the coupling parameter is negative, there always exists a purely imaginary unstable mode, indicating the instability of the black hole under consideration. When the coupling parameter is non-negative, such an instability disappears. On the other hand, with the increase of the non-negative coupling parameter, the violation of the SCC occurs at a larger critical charge ratio. Such an increase of the critical charge is nevertheless suppressed by the increase of the cosmological constant. In addition, we also find that different from the minimal coupling perturbation, the region for $\beta\ge1$ emerges in the near extremal black hole for the non-minimal coupling, where the increase of the cosmological constant can make such a region enlarged toward a smaller value of the coupling parameter. The existence of this region implies that the resulting curvature can continuously cross the $\mathcal{CH}$.

\section*{\bf Acknowledgements}

We are grateful to Hongbao Zhang for his helpful comments and suggestions. We also appreciate Zhiying Zhu for the helpful correspondence. This work is supported by the Natural Science Foundation of China under Grants No. 11705161 and Natural Science Foundation of Jiangsu Province under Grant No.BK20170481.

\bibliography{QNM}
\end{document}